\documentclass[prl,
				twocolumn,
				floatfix,
				showpacs,
				superscriptaddress,
				amsfonts,amsmath,amssymb]{revtex4}

\usepackage{upgreek,xspace}
\usepackage[]{graphicx}	

\newcommand{\mathcommand}[3][0]{\newcommand{#2}[#1]{\ensuremath{#3}}}
\mathcommand[1]{\ket}{\left| #1 \right\rangle}  
\mathcommand[1]{\bra}{\left\langle #1 \right|}  
\newcommand{\p}{$^{\text{31}}$P\xspace}
\newcommand{\sieight}{\ensuremath{{^{28}\text{Si}}}\xspace}
\newcommand{\sinat}{\ensuremath{{^\text{nat}\text{Si}}}\xspace}
\newcommand{\sinine}{\ensuremath{{^{29}\text{Si}}}\xspace}
\newcommand{\D}{\ensuremath{{D}^\text{0}}\xspace}
\newcommand{\DX}{\ensuremath{{D}^\text{0}{X}}\xspace}

\begin{document}

\title{Hyperfine structure and nuclear hyperpolarization observed in the bound exciton luminescence of Bi donors in natural Si}

\author{T.~Sekiguchi}
\author{M.~Steger}
\author{K.~Saeedi}
\author{M.~L.~W.\ Thewalt}
\email{thewalt@sfu.ca}
\affiliation{Department of Physics, Simon Fraser University, Burnaby, British
Columbia, Canada V5A 1S6}

\author{H. Riemann}
\author{N.~V. Abrosimov}
\author{N. N\"{o}tzel}
\affiliation{Institute for Crystal Growth (IKZ), 12489 Berlin, Germany}

\date{\today}

\begin{abstract}
As the deepest group-V donor in Si, Bi has by far the largest hyperfine interaction and also a large $I=9/2$ nuclear spin. At zero field this splits the donor ground state into states having total spin 5 and 4, which are fully resolved in the photoluminescence spectrum of Bi donor bound excitons. Under a magnetic field, the 60 expected allowed transitions cannot be individually resolved, but the effects of the nuclear spin distribution, $-9/2 \leq I_z \leq 9/2$, are clearly observed. A strong hyperpolarization of the nuclear spin towards $I_z = -9/2$ is observed to result from the nonresonant optical excitation. This is very similar to the recently reported optical hyperpolarization of P donors observed by EPR at higher magnetic fields. We introduce a new model to explain this effect, and predict that it may be very fast.
\end{abstract}


\maketitle
 
Recent proposals~\cite{Kane1998,Divincenzo2000,Vrijen2000,Stoneham2003,Morton2008} to use the electron and nuclear spins of shallow donor impurities as qubits for Si-based quantum computing (QC) have led to renewed interest in these systems~\cite{Morton2008,Karaiskaj2001,Yang2006,Vinh2008,McCamey2009,Yang2009a,Yang2009b}.  Most studies have focused on \p, the most common donor in Si, with an $I=1/2$ nuclear spin.  Most QC schemes involve enriched \sieight, as this eliminates the \sinine nuclear spin, but the removal of inhomogeneous isotope broadening~\cite{Karaiskaj2001} also enables an optical measurement of the donor electron and nuclear spin using the donor bound exciton (\DX) transition~\cite{Yang2006}, and furthermore allows for the hyperpolarization of both spin systems at very low magnetic fields by resonant optical pumping~\cite{Yang2009a}. McCamey et al.~\cite{McCamey2009} have reported a different effect in which P nuclear hyperpolarization can be achieved with nonresonant optical excitation in natural Si at high magnetic field and low temperature.

Bismuth is the deepest group-V donor in Si, with a binding energy of 70.98\,meV~\cite{Butler1975}, and is  monoisotopic ($^{209}$Bi), with a large $I = 9/2$ nuclear spin and a hyperfine interaction of 1475.4\,MHz, more than 12 times the 117.53\,MHz value for \p~\cite{Feher1959}. While invoked in some QC proposals~\cite{Stoneham2003}, Bi has not been the subject of recent study. It is interesting to note that the Bi \DX in Si is described in the earliest studies of bound excitons (BE) in semiconductors~\cite{Haynes1960,Dean1967} but has received little attention since then~\cite{Steiner1984}.  This likely resulted from the scarcity of samples and, until now, their low quality.  

Recently~\cite{Riemann2006}, Si:Bi samples have been grown from ultrapure natural Si (\sinat) using a floating-zone technique, for applications involving far-infrared lasers~\cite{Pavlov2002}.  Samples from those same crystals are studied here, and show very reproducible \DX no-phonon (NP) photoluminescence (PL) structure over a wide range of Bi concentration.  The spectra shown here are from a slice having a resistivity of 5.5\,$\Omega\cdot$cm, mostly due to Bi, since the residual B and P concentrations are estimated to be at least an order of magnitude less than the Bi concentration.  The sample was mounted without strain in a high homogeneity (0.01\,\%) split pair superconducting magnet dewar in Voigt configuration, with magnetic field $B$ approximately along [100], and immersed either in liquid He or cold He gas. Above-gap excitation of 400\,mW was provided by a 1047\,nm laser, and the collected PL was analyzed by a Bomem DA8 interferometer at a resolution of 1.8$\,\mu$eV full width at half maximum (FWHM). No effort was made to control either the excitation or the PL polarization. 

\begin{figure}[t] \centering
\includegraphics[width=.85\columnwidth]{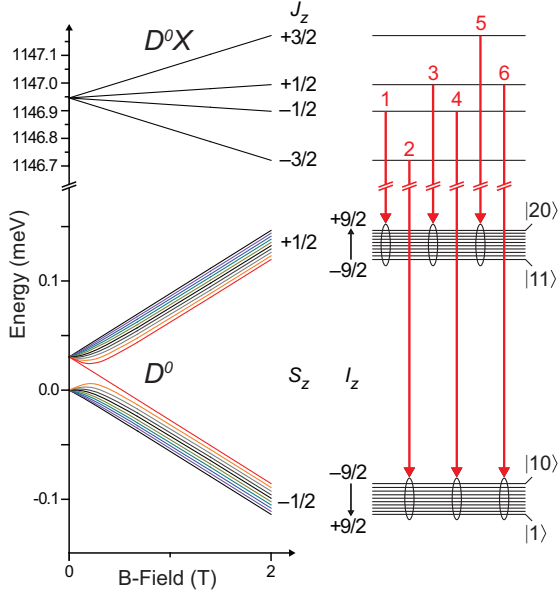}
\caption{\label{fig:leveldiagram}
(Color online) Zeeman level diagrams for Bi \D and Bi \DX indicating the origin of the PL structure from 0 to 2\,T. $J_z$ labels the hole angular momentum projection in \DX, while $S_z$ and $I_z$ label electron and nuclear spin projections, respectively, in \D. The energy scale for \DX is compressed by a factor of 4. On the right the allowed transitions at 2\,T are labelled from 1 to 6 in order of increasing energy, with each having 10 hyperfine subcomponents.
}
\end{figure}

A level scheme indicating the expected PL Zeeman transitions between \DX and the donor ground state (\D) is shown in Fig~\ref{fig:leveldiagram}. While this level scheme is similar to one introduced to explain the P \DX hyperfine structure in \sieight~\cite{Yang2007,Yang2009a}, it must be stressed that the P spectra were collected using PL excitation (PLE) spectroscopy, in which the thermalizing initial states of the transitions were the \D states, whereas the present data is collected via PL spectroscopy, where the initial states are the \DX states, resulting in quite different thermalization behaviour. At zero magnetic field, the \D ground state is split into a doublet having total spins 5 and 4, and separated by 5 times the hyperfine interaction, or 7377\,MHz (30.51$\,\mu$eV)~\cite{Feher1959}. For nonzero $B$ the \D states split according to the projection of the electron spin, $S_z$, into two main branches, each of which has ten hyperfine subcomponents.  The donor hyperfine levels are labelled from \ket{1} to \ket{20} in order of increasing energy (this ordering only changes at extremely high field).  As for all substitutional donor \DX in Si~\cite{Thewalt1982}, in the ground state the two electrons must occupy the doubly-degenerate $1s \Gamma_1$ shell, forming a spin zero singlet, so the \DX splitting is determined only by the projection of the hole spin, $J_z$ (the nuclear Zeeman energy is small at these fields and is ignored, and the hyperfine coupling with the p-like hole is negligible).  The six dipole-allowed \DX PL transitions are labelled from 1 to 6 in order of increasing energy, and each can be split into ten subcomponents by the hyperfine interaction. 

\begin{figure}[t]
  \centering \includegraphics[width=.8\columnwidth]{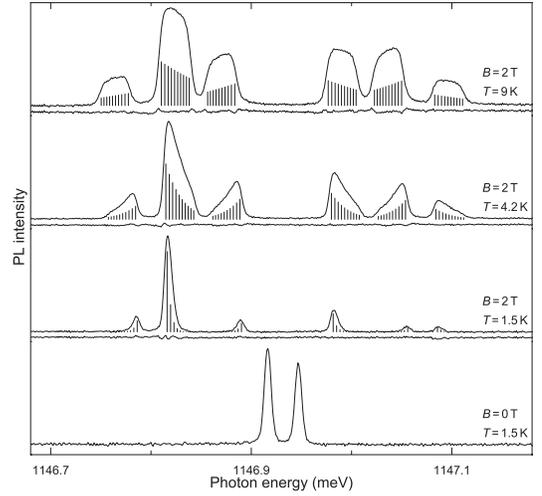}
  \caption{\label{fig:fits}
PL of the Si:Bi NP \DX transitions is shown at zero field, and at 2\,T for 3 different temperatures. Vertical lines indicate relative strengths of individual hyperfine components obtained from the fit. The residual error of each fit is shown just below the spectrum.
}
\end{figure}

The zero-field Si:Bi hyperfine doublet can be completely resolved in all of our samples, as shown at the bottom of Fig.~\ref{fig:fits}.  The observed splitting is equal to the value expected from ESR~\cite{Feher1959}, and the relative intensities are in good agreement with the degeneracies of the two states. The linewidths of 7.9$\,\mu$eV FWHM are slightly larger than the 5.7$\,\mu$eV width of the P \DX in \sinat. Given that this linewidth is also larger than the Bi hyperfine interaction, it is not surprising that individual hyperfine components cannot be resolved when a field is applied, although all 60 hyperfine components are expected to be completely resolved in even a moderately enriched \sieight:Bi sample. The hyperfine splittings are nevertheless evident in the Zeeman spectra even in \sinat, given the energy spread of the ten components spanning $-9/2 \leq I_z \leq 9/2$, as can be seen in the three spectra in Fig.~\ref{fig:fits} taken at $B = 2$\,T, the lowest field at which the six allowed PL transitions have no overlap.  

The top spectrum is at a temperature of $T=9$\,K, as estimated from the width of the free exciton (FE) PL, and the six allowed transitions appear roughly rectangular, as expected if the $-9/2 \leq I_z \leq 9/2$ states are equally populated (the total hyperfine energy difference between the $-9/2$ and $9/2$ sublevels is much less than $k_BT$ at any temperature used here).  The observed spectra have been fit by calculating the transition energy of each of the sixty unresolved hyperfine components, and adding a line having a shape similar to the zero-field lineshape (but somewhat narrower) at that energy, with a relative intensity determined by \DX $\rightarrow$ \D selection rules, initial state thermalization, and the nuclear polarization.  The known electron $g$-factor is used in the fits, and the \DX hole $g$-factors are adjustable parameters, together with the diamagnetic shifts~\cite{Kaminskii1980} and the temperature dependence of the band gap energy, all of which are optimized across the spectra at various temperatures and $B$ fields.  An effective temperature, slightly higher than the nominal temperature, is obtained from matching the thermalization between the six main \DX transitions.  The residual error of the fit is shown under each of the three spectra.

A nuclear polarization term has to be included to explain the skew of the observed spectra, especially at lower $T$, and given that the individual hyperfine components cannot be resolved, we make the simple assumption that the polarization per step of $\Delta I_z = 1$, $P_\text{fit} = [N(I_z+1)-N(I_z)]/[N(I_z+1)+N(I_z)]$, is independent of $I_z$. The relative intensities and energies of the hyperfine subcomponents are indicated by the vertical lines under the PL components. It is clear that large nuclear polarizations are being produced, particularly at low $T$, where a substantial fraction of the nuclei are being polarized into the $I_z = -9/2$ state.  The nuclear polarization results of the fits at $B = 2$\,T, as well as at $T = 1.5$\,K and $B = 6$\,T are summarized in Table~\ref{tab:pol}. These nuclear hyperpolarizations under nonresonant optical excitation, in a direction opposite to what would be expected for the equilibrium nuclear polarization of the $S_z = -1/2$ branch, are very similar to the effect recently reported for \sinat:P at 8\,T~\cite{McCamey2009}.

\begin{table}[t]
\caption{\label{tab:pol}Results of the PL fitting procedure for several $B$-fields and nominal $T$ ($T_\text{bath}$). $T_\text{fit}$ is the temperature determined by the observed \DX thermalization, $P_\text{fit}$ is the nuclear polarization per step of $\Delta I_z = -1$, $P_{e}$ is the expected equilibrium polarization of \D electrons, and $N(-9/2)$ is the fraction of Bi donors hyperpolarized into the nuclear spin state $I_z = -9/2$.}
\begin{ruledtabular}
\begin{tabular}{rrrrrr}

   $B$ (T) & $T_\text{bath}$ (K) & $T_\text{fit}$ (K) & $P_\text{fit}$ (\%) & $P_{e}$ (\%) &   $N(-9/2)$(\%) \\
\hline
         2 &        8.8 &        8.9 &    $-3(1)$ &      $-15$ &      13(1) \\

         2 &        4.2 &        4.7 &   $-10(2)$ &      $-27$ &      21(2) \\

         2 &        1.5 &        1.7 &  $-54(15)$ &      $-65$ &     69(13) \\

         6 &        1.5 &        1.5 &  $-79(21)$ &      $-99$ &     86(14) \\

\end{tabular}  
\end{ruledtabular}
\end{table}

\begin{figure}[t]
 \centering
  \includegraphics[width=.8\columnwidth]{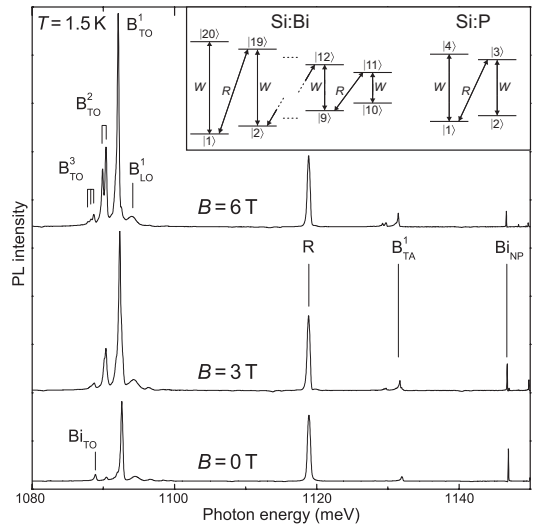}
  \caption{\label{fig:pol_spec}
PL spectra of the \sinat:Bi sample at low $T$ and fields of 0, 3 and 6\,T.  The excitation conditions and the intensity scale are the same for all three spectra. The inset shows a simple labelling scheme of the Bi and P hyperfine states for a discussion of the origin of the nuclear hyperpolarization.
}
\end{figure}

McCamey et al.~\cite{McCamey2009} explained this effect in terms of an Overhauser-like process, which we summarize with reference to the inset in Fig.~\ref{fig:pol_spec}. The basis of their model is that the P electron spin thermalizes via the $W$ processes at a temperature at or near the actual sample temperature $T$, whereas the $R$ process equilibrates the \ket{1} and \ket{3} states at some higher temperature $T^*$. At low $T$ and high $B$, where the equilibrium electron spin polarization is large, this will pump the \D population preferentially into \ket{2}.  While this process may certainly play a role in the observed optical polarization, it is difficult to see how one could arrive at \emph{ab-initio} estimates of the two temperatures, other than by using them to fit the observed polarization.

We propose a different effect, inherent in the capture of FE to form \DX for all substitutional \D in Si, when $B$ is high enough and $T$ is low enough to generate significant electron polarization.  For substitutional donors in Si, the \DX ground state has two electrons with antiparallel spins. At high field, both the \D electron and the FE electron will be well polarized into $S_z = -1/2$, and the formation of \DX in its ground state requires the flipping of one of these electron spins, which remains energetically favourable since the \DX localization energy at the fields in question is still considerably larger than the electron Zeeman splitting. \DX having two $S_z = -1/2$ electrons might be an intermediate state in the formation of the ground state \DX, but this requires the captured electron to occupy the barely bound $1s\Gamma_{3,5}$ valley-orbit excited states, which lies $\sim$4.3\,meV ($\sim$6.9\,meV) above the ground state for P (Bi)~\cite{Thewalt1982}. At $T \leq 15$\,K these excited states are efficiently thermalized to the ground state within the \DX lifetime. In any case, formation of the ground state \DX requires an electron spin flip ($\Delta S_z = 1$) which could occur via the spin-orbit effect.

However, substitutional donors have another mechanism for achieving this electron spin flip during the capture of polarized FE onto polarized \D. For P at high $B$ and low $T$ and without optical excitation, states \ket{1} and \ket{2} will each have nearly 50\,\% of the total population (the energy difference between them being much less than $k_BT$). Note that for P, states \ket{2} and \ket{4} are pure \ket{S_z, I_z} states, whereas states \ket{1} and \ket{3}, while tending towards \ket{-1/2, 1/2} and \ket{1/2, -1/2}, respectively, at high field, always have an admixture of the other component.  \D in state \ket{2} can only capture polarized FE via spin-orbit flipping of one of the electron spins. Those in state \ket{1} can use the admixture of \ket{1/2, -1/2} to flip the electron spin and form the \DX, while also flipping the nuclear spin for a total spin change of 0, and driving the nuclear population into $I_z = -1/2$, as is observed~\cite{McCamey2009}. It is unlikely that the hole spin changes during the FE capture, but in any case the hole in the FE is already polarized into the $J_z = -3/2$ state, so this would only increase the change in angular momentum. The same process will apply to Si:Bi, as shown on the left of the inset in Fig.~\ref{fig:pol_spec}. For Bi, only the \ket{10} and \ket{20} states are not mixed, and capture of polarized FE by \D in states \ket{1} through \ket{9} via this flip-flop process decreases $I_z$ by 1, leading to a buildup of population in $I_z = -9/2$. 

This model has testable consequences, namely, when the \D population builds up in the most favoured state (\ket{2} for Si:P and \ket{10} for Si:Bi), capture of polarized FE to form \DX via the flip-flop process is no longer possible. In Fig.~\ref{fig:pol_spec} we show PL spectra of our \sinat:Bi sample on a wider energy scale at fields of 0, 3, and 6\,T, all at low $T$.  The zero-field spectrum is dominated by the no-phonon line of the Bi \DX (Bi$_\text{NP}$) and the transverse optical (TO) phonon replica of the boron acceptor BE (B$^1_\text{TO}$). The optical phonon Raman line of the 1047\,nm excitation laser is also observed.  Even though the Bi concentration is much higher than that of B, the B PL is stronger, since the B BE has higher radiative quantum efficiency than the Bi \DX which has a very short Auger lifetime~\cite{Steiner1984}.  Note that B$^2_\text{TO}$, the two-exciton B bound multiexciton complex (BMEC) is very weak at zero field, as expected for a sample containing $\sim 10^{15}$cm$^{-3}$ Bi, since the Bi capture most of the FE, keeping the FE concentration low.  At 3\,T the absolute Bi$_\text{NP}$ intensity has decreased, and the B$^1_\text{TO}$ intensity increased, but even more noticeable is the large increase in the B$^2_\text{TO}$ intensity, and the appearance of B$^3_\text{TO}$, indicating a large increase in the FE density (at zero field the lowest energy feature is the TO replica of the Bi \DX (Bi$_\text{TO}$), but at higher fields the B$^3_\text{TO}$ dominates).  These changes are even more pronounced at 6\,T.  Between 0 and 6\,T the intensity of the Bi \DX PL relative to the B-related PL has decreased by a factor of 9. Almost exactly the same changes in relative intensity were observed between B and the P \DX in a \sinat sample containing only B and P.

This supports our model for the origin of the optically induced nuclear hyperpolarization, since once the Bi donors are hyperpolarized into the \ket{10} state they can no longer capture polarized FE via this process, and hence the Bi PL decreases and the FE density increases, causing the B BE and BMEC lines to increase in intensity (the electron(s) in the B BE and BMEC can all have $S_z = -1/2$). Note that the lowest Zeeman state of the B BE and the \DX both have a single dipole allowed PL transition~\cite{Thewalt1982}, so the change in the relative PL intensities does not result from selection rules. The fact that B$^1$ has a single line at high $B$ and low $T$ while B$^2$ has two and B$^3$ three is well understood~\cite{Thewalt1982}. The large increase in FE density made evident by the growth of the B BMEC lines at high field does not result from the fact that the FE are being polarized into a 'dark' state with $S_z = -1/2$ and $J_z = -3/2$, since even at zero field the FE lifetime is completely dominated by capture onto donors and acceptors to form BE, a process much faster than the ms-long FE radiative recombination.

We have been unable to observe transient polarization effects, leading us to suspect that the nuclear hyperpolarization might be too fast under the excitation conditions needed to observe PL for our low speed detector to follow. This surprising and potentially important possibility can be understood in terms of our model. Capture of a single polarized FE onto a \D in a mixed state should flip its nuclear spin with high probability, and under our excitation conditions this capture time is likely in the 10 to 100\,$\mu$s range. The $\sim$100\,s polarization time observed by McCamey et al.~\cite{McCamey2009} could have resulted from much lower levels of optical excitation used in their experiment.

In conclusion, we have resolved the zero-field hyperfine splitting of the Bi donor in the PL spectrum of the NP line of the \DX transition in \sinat, and have observed a large nonresonant optical hyperpolarization of the nuclear spin under moderate to large $B$ field, which is very similar to results recently reported for P~\cite{McCamey2009}. At 6\,T and 1.5\,K, essentially all of the donors are pumped to the $I_z = -9/2$ state. We propose a new model for the origin of this hyperpolarization, which is supported by changes in the overall PL spectra with magnetic field. Further experiments are needed to test the prediction that this hyperpolarization mechanism might in fact be extremely fast. PL in \sieight:Bi would clearly resolve all 60 \DX hyperfine components, and absorption or PLE measurements on the Bi \DX should result in a more direct measurement of the populations in all of the \D states. 

Acknowledgements: We thank T.\ D.\ Ladd for useful discussions. This work was supported by the Natural Sciences and Engineering Research Council of Canada (NSERC).


\end{document}